\documentclass{iaus}
\usepackage{graphicx}

\title[Analysis Techniques for SBs] 
{Modern Analysis Techniques for Spectroscopic Binaries}

\author[H. Hensberge \& K. Pavlovski]  
{Herman Hensberge$^1$ \and Kre\v{s}imir Pavlovski$^2$}

\affiliation{$^1$Royal Observatory of Belgium, Ringlaan 3, B-1180 Brussels, Belgium 
             \break email: herman.hensberge@oma.be\\[\affilskip]
$^2$Department of Physics, University of Zagreb, Bijeni\v{c}ka 32, HR-10000 
Zagreb, Croatia   
\break email: pavlovski@phy.hr}

\pubyear{2006}
\volume{240} 
\pagerange{?--?}
\date{?? and in revised form ??}
\jname{Binary Stars as Critical Tools and Tests in Contemporary Astrophysics}
\editors{W. Hartkopf, E. Guinan \& P. Harmanec, eds.}

\begin{document}

\maketitle

\begin{abstract}
Techniques to extract information from spectra of unresolved multi-component
systems are revised, with emphasis on recent developments and practical
aspects. We review the cross-correlation techniques developed to deal with 
such spectra, discuss the determination of the broadening function and 
compare techniques to reconstruct component spectra. The recent results 
obtained by separating or disentangling the component spectra is summarised. 
An evaluation is made of possible indeterminacies and random and systematic 
errors in the component spectra.  
\end{abstract}

\firstsection 

\section{Introduction}\label{sec:intro}

We summarize recent progress and discuss relations between analysis techniques 
to determine the orbital parameters and the intrinsic spectra of components in 
multiple systems. Progress in applying these techniques has been driven by 
very different astrophysical applications. The improvement of templates and 
the increase in sensitivity in cross-correlation techniques has been driven 
by planet search programmes. The emphasis is thus on the rich  line spectra 
of cool stars. Broadening functions are of general interest, but at present 
applications are restricted to very-short period close binaries, where the 
rotational broadening hampers the detection and the analysis of the components.
Techniques to separate and disentangle the spectra of the components from 
observed multi-component spectra have served mainly in the area of detached, 
hotter stars. 

The emphasis in this contribution is put on practical issues, with the aim 
to guide potential users of these techniques to obtain in the most efficient 
way appropriate observation sets and to constrain the transfer of systematic 
errors to the output quantities. Excellent reviews on the different techniques 
used for reconstructing the component spectra are available in \cite{Gies2004}, 
\cite{Hadrava2004} and \cite{Holmgren2004}. The use of broadening functions 
has been summarized in \cite{Rucinski2002} 
and practical aspects are summarized in \cite{Rucinski1998}.  
An overview of 1D and 2D cross-correlation, with several examples, is found  
in \cite{Hilditch2001}.

This paper is structured as follows: Section 2 deals with improvements in  
measuring the orbital movement with cross-correlation techniques. Section 3 
discusses the determination of broadening functions. Section 4 summarizes 
the progress in reconstructing component spectra since the Dubrovnik meeting 
in 2003 (\cite{Hilditchetal2004}). Section 5 addresses the risk of 
introducing spurious patterns in the intrinsic component spectra.  

\section{Cross-correlation techniques}\label{sec:ccf}

\subsection{Templates}\label{sub:ccft}

Cross-correlation of the stellar spectrum with a template spectrum supposed 
to represent well the intrinsic stellar one is already several decades a 
standard technique to measure Doppler shifts. Obviously, a single-lined 
template cannot represent well multiple component spectra with intrinsically 
different components. Therefore, the need arises to cross-correlate with a 
template involving two or more components, each with a time-dependent shift 
of its own, 
$$t(x;s_1,\ldots,s_n) = t_1(x-s_1) + \sum_{i=2}^n \alpha_{i1} t_i(x-s_i) 
\eqno(1)
$$ 
This does not increase the computing time enormously, as it has been shown 
(\cite{ZuckerMazeh1994}; \cite{Zuckeretal1995}) that the $n$-dimensional 
surface of the cross-correlation function ({\sc ccf}) can be reconstructed from 
a non-linear combination of 
$\frac{n(n+1)}{2}$ one-dimensional {\sc ccf}s, 
namely the {\sc ccf}s computed for  all pairs $(t_i,t_j)$, $i>j$, and  
$(t_i, \mathrm{stellar \,\, spectrum})$. The relative strength parameters 
$\alpha_{i1}$ can be fixed by external conditions or optimized during the 
cross-correlation process. In the latter case, these parameters can be 
computed analytically as function of the one-dimensional {\sc ccf}s. The 
technique has been applied successfully for $n = 2$ and $n = 3$. An example 
of a two-dimensional cross-correlation surface can be found in 
\cite{Zucker2004}.

\subsection{{\sc ccf} sensitivity}\label{sub:ccfw}

The need to detect faint components, or analyse low signal-to-noise data, 
often obtained with echelle spectrographs that impose on the data a strong 
modulation in each spectral order, has prompted investigations to find out 
how to improve the sensitivity level of the cross-correlation technique under 
the assumption that random noise is the dominating source of error. 

\cite{Bouchyetal2001} noted that bins at large spectral gradients 
and high signal-to-noise level contain most velocity information. They 
propose a weighting scheme, very similar to what is done in optimal extraction 
techniques in {\sc ccd} spectroscopy, to obtain a {\sc ccf} with minimum 
variance. Their paper includes also a discussion of the radial-velocity 
information content intrinsically present in the data as a function of 
the wavelength range, the spectral type (F to K), the rotational broadening 
and the spectrograph resolution. 

\cite{Zucker2003} used maximum likelihood principles to argue against a 
linear addition of {\sc ccf}s obtained in different spectral regions. He 
gives a non-linear combination formula, which reduces in the limit 
of low signal-to-noise to a quadratic average of the {\sc ccf}s. 

Finally, \cite{Chelli2000} argues for an alternative algorithm to determine 
Doppler shifts. It is based on a rigorous approach in the spectral Fourier 
domain that uses a weighted analysis of the cross spectrum phase between the 
high resolution spectra of the object and an appropriate template. In 
\cite{Gallandetal2005} it is applied to stars of spectral type A and F. 

In relation to earlier spectral types, \cite{Griffinetal2000} investigated the 
impact of spectral mismatch in the B8--F7 spectral-type range on the accuracy 
of radial-velocity measurements and suggest a suitable window around 
$\lambda 4570$, though with fast rotation a very large window may be more 
appropriate to overcome the lower intrinsic radial velocity content. 

\subsection{Single-lined spectroscopic binaries}\label{sub:ccf1}

Very recently, \cite{ZuckerMazeh2006} presented a method to derive in a 
self-consistent way radial-velocity changes for a set of spectra of a 
single-lined binary, by considering the whole matrix of {\sc ccf}s computed 
for all pairs of input spectra. In this case, an external template is not 
needed. It is shown to be equivalent to the use of a properly-weighted 
average of all input spectra (after alignment to compensate for the Doppler 
shifts) as a template. This technique goes strongly in the direction of the 
disentangling of spectra applied to a one-component system, the latter 
prefering to optimize for orbital parameters rather than radial velocity 
shifts in order to increase the robustness of the technique. 

\section{Broadening function}\label{sec:bf}

Despite its wide-spread use, the cross-correlation technique has few 
disadvantages which are particular relevant in the case of multiple systems. 
The shape of the {\sc ccf} depends on the shape of the spectrum, because 
chance overlaps of different spectral lines in the stellar spectrum and the 
shifted template contribute to a fringing pattern in the {\sc ccf}. In the 
case of the multiple-peaked {\sc ccf} corresponding to a multiple system, this 
pattern may lead to biased radial velocities. 
In addition, the peaks in the {\sc ccf} are wider than the spectral lines in 
the stellar spectrum, because the width of the spectral lines in the template 
adds to the resulting width of the {\sc ccf}. 

Both disadvantages are avoided when solving for a broadening function 
({\sc bf}) that, convolved with the template, represents the stellar spectrum. 
The {\sc bf} then contains only the {\it additional} broadening mechanisms 
affecting the star (and not the template) and, possibly, reflects also 
time-dependent instrumental effects. As a consequence, different components 
separate easier in the {\sc bf} than in the {\sc ccf}. This 
method is developed by \cite{Rucinski1992} and applied to close 
binaries with orbital periods less than one day in a series of papers 
(see \cite{Pribullaetal2006} and \cite{RucinskiDuerbeck2006} for recent papers 
on northern and southern stars). 

The position of the {\sc bf} reflects the radial velocity. In 
\cite{Pribullaetal2006} the position is measured fitting  rotational 
profiles rather than Gaussian profiles. The integrated intensity of the 
different components in the {\sc bf} is directly proportional to the light 
ratios, in the case of identical line blocking coefficients and on condition 
that the continuum is determined correctly. The latter is not trivial in view 
of the large rotational broadening in the rich line spectra. As noted 
various times by Rucinski, proper modelling of the {\sc bf} extends its 
usefulness to studies involving stellar spots, limb darkening, stellar shape 
and other factors contributing to the broadening of spectral lines and its 
variation with orbital phase. The full exploitation of the technique lies
clearly still in the future. 

Technically, the determination of the {\sc bf} reduces to a linear problem 
that is suitably overdetermined when the stretch of spectrum analyzed is 
much longer than the width over which the {\sc bf} must be solved. The latter 
is of the order of the sum of the Doppler separations due to the orbital 
motion plus the width of the corresponding {\sc bf}s.

\section{Reconstruction of component spectra}\label{sec:comp}

\subsection{Separation and disentangling techniques}\label{sub:tech}

When observed spectra are the sum of intrinsically time-invariant components 
that shift with respect to each other, depending on the time of observation, 
then the intrinsic component spectra can be reconstructed from a time-series 
of observed spectra by exploiting the relative Doppler shifts. The weight of 
a component may vary with time, but - at least in the original formulation - 
not its intrinsic spectrum. This excludes the use of spectra obtained in 
partial eclipses and systems with a component showing variations in 
spectral-line shape. 

Early attempts to separate the spectra of the stars in composite spectra date  
back at least to \cite{Wright1952}. A series of papers was initiated by 
\cite{Griffin1986} for systems consisting of a cool giant and a hotter 
main-sequence star. They searched for a suitable template for the giant 
spectrum and reconstructed the hotter component by subtracting from the 
observed spectra the cooler template in the right amount, and properly shifted.
As shown in \cite{Griffin2002}, the method fails when the cool giant is 
peculiar. A technique not based on assumptions about the shape of one of the 
component spectra is needed in such cases. 

Several such methods were proposed in the last decennium of the 20th century.
They require that the number of components is specified a priori. 
\cite{BagnuoloGies1991} introduced a tomographic technique to separate the 
component spectra once the mutual Doppler shifts are known. They propose an  
iterated least-squares technique ({\sc ilst}) as solution scheme. 
\cite{SimonSturm1994} formulated a solution for the more complex 
problem to separate the spectra of the components and to determine 
self-consistently the orbital parameters in an iteration scheme during which 
orbital parameters and component spectra are improved in turn. One refers 
to this more complex problem as the disentangling of the component spectra. 
Orbital parameters are optimised by a $\chi^2$-type minimisation of the 
residuals between the observed spectra and their reconstructed model, while 
the problem is linear in the relative intensities of the component spectra. 
Solving for the latter unknowns involves a large set of overdetermined, but 
rank-deficient matrix equations (number of spectral bins times number of 
observed spectra) with a large number of unknowns (somewhat larger than the 
number of spectral bins times the number of components in the spectrum). This 
is performed by use of the singular value decomposition technique. 
The computational requirements were reduced significantly when 
\cite{Hadrava1995} showed that in the space of the Fourier components of the 
spectra, the huge number of coupled equations reduces to many small sets of 
equations ($\frac{n_{\mathrm{bins}}}{2} + 1$ 
% HH ($n_{\mathrm{bins}}/2 + 1 $
independent sets of $n_{\mathrm{comp}}$ complex equations), 
each set corresponding to one Fourier mode.  

Recently, \cite{GonzalezLevato2006} developed a method used 
earlier by \cite{Marchenkoetal1998}. They use an iterative scheme, using 
alternately the spectrum of one component to predict the spectrum of the 
other one. In each step, the calculated spectrum of one star is used to remove 
its spectral features from the observed spectra and then the resulting 
single-lined spectra are used to measure the Doppler shifts for the remaining 
component and to compute its spectrum by an appropriately shifted combination 
of the single-lined spectra. This is a tomography-like method with iterations 
on the Doppler shifts. 

In principle, solving the problem in velocity space or in Fourier space is 
equivalent, but there are some practical differences (see also 
\cite{Ilijicetal2001}). One aspect relates to the edges of the considered 
spectral regions where, depending on the orbital phase, information on 
particular bins in the intrinsic spectra enters and leaves the selected 
spectral range in the observed spectra. \cite{SimonSturm1994} solve for the 
component spectra over a spectral range slightly larger than in the observed 
spectra, although not all input spectra carry information on the outer bins. 
On the contrary, in Fourier space, the spectra are considered to be periodic 
and data are wrapped around. Another aspect relates to sampling non-integral 
bin velocity shifts. One can use interpolation schemes on the original grid 
or oversample the spectra in finer velocity grids. In pathological cases 
(strong lines at the edges of the selected interval, singular equations, ...) 
the result may be significantly different.  

More important is the difference in weighting options: in velocity space, 
each bin can be weighted proportional to its precision, and blemished or 
useless data can be masked out (e.g. non-linear pixels, interstellar lines, 
telluric lines, \ldots). Alternatively, the Fourier modes can be weighted 
allowing  e.g.\ to diminish the impact of low-frequency Fourier components in 
the optimalisation process; it turns out to be easier in Fourier space to 
control and remedy the occurrence of spurious patterns in the component 
spectra due to numerical singularities or bias in the observed spectra. 
A combination of both techniques is an option: exploit the computational speed 
of the Fourier analysis to find the orbital solution, and separate the 
component spectra with known orbital parameters in velocity space to allow for 
the proper masking of the data. Both methods react also different on certain 
types of bias in the input data (\cite{Ilijic2004}; \cite{Torresetal2007}).

\subsection{Input data}\label{sub:inp}

The observed spectra must be sampled in velocity bins (logarithm of 
wavelength). In order to avoid resampling noise, and since the resolution of 
echelle spectra is often in good approximation proportional to velocity 
and not to wavelength, such sampling is best performed immediately when 
reducing the raw data. Ideally, any resampling during the iterative 
reconstruction process should start again from non-resampled data. 

The techniques described here are differential in the sense that they rely 
on the time-variability of the Doppler shifts between pairs of components, 
and thus deliver differential velocities -- the systemic velocity has to be 
determined separately and involves the identification of spectral lines. 
An ideal data set covers fairly homogeneously all relative Doppler shifts and 
does not concentrate on spectra observed near maximum line separation. 
In eccentric orbits, a fairly small range of orbital phases near periastron 
is suitable to cover all velocity shifts, although a good orbit determination 
may require a better phase coverage. 

Spectra in mid-eclipse are extremely useful to stabilise the low-frequency 
components in the output spectra (see also Section \ref{sec:risk}). They also  
allow to circumvent the indeterminacy in the level of line blocking in the 
intrinsic component spectra (or, in other words, in their zero-point level) 
when the light ratio of the components is time-independent. These light ratios 
can be determined spectroscopically during the reconstruction process, or may 
be fixed by external conditions, as e.g.\ high-precision photometry, depending 
on which choice provides the most precise information. The relative light 
contributions or, equivalently, the line blocking in the component spectra, 
can also be estimated accurately from the observed spectra when the components 
have very deep absorption lines, since no spectral line in the intrinsic 
component spectra should cross the zero-intensity level. 
Hence, in  absence of eclipses this calls for observation of spectral regions 
with deep absorption lines, which is especially feasible in slowly rotating 
cooler stars. In absence of this fortunate situation, light ratios can be 
bracketed in a more indirect way, e.g.\ by requiring that components 
have identical abundances (if realistic), or by requiring that faint and 
strong lines of the same ion should give the same abundance, or by bracketing 
the strength of specific absorption lines, etc. Fortunately, in the case of a 
constant light ratio between all components, the disentangling process can be 
separated from the decision which light ratio to apply 
(e.g.\ \cite{Ilijicetal2004}). 

The random noise in the output spectra is reduced by the combination of  
$n_{\mathrm obs}$ input spectra, but increases inversely proportional with 
the relative light contribution $\ell_j$ of each component $j$. A useful, but 
somewhat optimistic signal-to-noise estimate may be obtained from 
$$ {\mathrm (S/N)}_j = {\mathrm (S/N)}_{\mathrm obs} 
n^{\frac{1}{2}}_{\mathrm obs} \ell_j 
\eqno(2)
$$
Hence, the spectrum of the dominant component is often less noisy than the 
observed spectra, but a large set of input spectra is needed to obtain a 
high-quality spectrum of a faint component. With less random noise in the 
output spectra, systematic noise in the observed spectra may become the 
dominant source of uncertainty in the component spectra (Sect. \ref{sec:risk}).

\subsection{Application domain}\label{sub:appl}

\begin{table}
\caption{\label{tab:lit} Recent applications of separating or disentangling
 of component spectra. Codes: {\sc ft} Fourier analysis, 
 {\sc idd} iterative Doppler differencing, {\sc ilst} tomography, 
 {\sc svd} velocity space analysis, {\sc nlls} non-linear least-squares}
\begin{tabular}{lllll}
\noalign{\smallskip}\hline\noalign{\smallskip}
source & code & target(s) & comment \\
\noalign{\smallskip}\hline\noalign{\smallskip}
Budovi\v{c}ov\'{a} \etal\ 2004 & FT & o\,And & Be, 3 comp. disent. ; orbit \\
Harmanec \etal\ 2004 & FT & $\kappa$\,Sco & NRP $\beta$\,Cep \\  
Zwahlen et al.\ 2004 & FT & Atlas  & distance to Pleiades \\
Fr\'emat \etal\ 2005 & FT & DG\,Leo & (Am+Am)+A8 $\delta$\,Sct; abund. \\
Hilditch \etal\ 2005 & SVD/NLLS & SMC & 40 OB-type EBs, fund. par. \\
Lehmann \& Hadrava 2005 & FT & 55\,UMa & triple, fund. p., 1300 sp. \\ 
Ribas \etal\ 2005 & FT & EB in M31 & fund. p. ({\sc todcor} + separ.) \\
Pavlovski \& Hensberge 2005 & FT & V578\,Mon &  abund. early-B, NGC\,2244 \\
Saad \etal\ 2005 & FT & $\kappa$\,Dra & Be, emiss.; sec. undetected \\
Uytterhoeven \etal\ 2005 & FT & $\kappa$\,Sco & NRP $\beta$\,Cep, 700 sp. \\ 
Ausseloos \etal\ 2006 & FT & $\beta$\,Cen & NRP $\beta$\,Cep, fund. p., 400 sp. \\ 
Bak\i \c{s} \etal\ 2006 & FT & $\delta$\,Lib & Algol-type \\
Boyajian et al.\ 2006 & ILST & HD\,1383  & B0.5Ib+B0.5Ib, fund. p.  \\ 
De Becker \etal\ 2006 & IDD & HD\,15558 & IC\,1805, detection sec. O7V  \\ 
Gonz\'{a}lez \& Levato 2006 & IDD & HD\,143511 & fund. p., ecl. from sp., BpSi \\
Gonz\'{a}lez \etal\ 2006 & IDD & AO\,Vel & quadruple, BpSi primary \\
Hensberge \etal\ 2006 & FT & RV\,Crt & fund. p., pre-MS \\
Hillwig et al.\ 2006 & ILST & Cas\,OB6   & 13 O-type stars, fund. p. \\ 
Hubrig \etal\ 2006 & IDD & AR\,Aur & line shape var. B9(HgMn)  \\
Koubsk\'{y} \etal\ 2006 & FT & HD\,208905 & Cep\,OB2, triple \\
Linnell \etal\ 2006 & FT & V360\,Lac & crit. rot. Be, fund. p. \\ 
Martins \etal\ 2006 & IDD & GCIRS16SW & Gal. Center, He{\sc i}\,2.1$\mu$m \\
Pavlovski \etal\ 2006 & FT/SVD & V453\,Cyg & He abundance \\  
Chadima \etal\ 2007 & FT & $\beta$\,Lyr & distorted star\,+\,accr. disk \\ 
Lampens \etal. 2007 & FT & $\theta^2$\,Tau & $\delta$\,Sct in Hyades; orbit \\ 
Pavlovski \& Tamajo 2007 & FT & CW\,Cep, V478\,Cyg & He abundance \\
\noalign{\smallskip}\hline
\end{tabular} \\
\end{table}

Recent applications, published after the reviews of \cite{Gies2004} and 
\cite{Holmgren2004}, are given in Table~\ref{tab:lit}. The columns give the 
names of the authors, the algorithm code used (acronyms as in \cite{Gies2004} 
and {\sc idd} = iterative Doppler differencing for the Gonz\'{a}lez \& Levato 
method), the target name and comments. The applications cover a wide range 
of spectral types (O to G) in binaries, spectroscopic triple systems and, 
in a single case, a spectroscopically quadruple system. Many of these 
systems proved intractable with classical techniques. Some of the components 
contribute less than 10\% to the total light. In various applications, the 
data are combined with photometric and/or astrometric data. Some applications 
involve an impressive amount of several hundreds to more than one thousand 
spectra. Often, short spectral intervals are used, because they serve the 
purpose, but in other works large pieces of spectrum are successfully 
reconstructed. 

These studies have led to the determination of flux ratios, the detection 
of eclipses, the spectroscopic detection of components, the analysis of 
the atmospheric parameters as for single stars, including (peculiar) 
abundances, the assignment of line profile variability to specific components 
and their study free of the diluting effects of other components, the 
detection of changes in a close-binary orbit caused by the tidal interaction 
of a third companion, and the determination of stellar masses and distances 
(the latter from the Pleiades to the Local Group galaxies). 

Among the high {\sc s/n} and high-resolution applications, several scientific 
programmes aim to study the chemical composition of the atmosphere: an 
observational study of rotational mixing during the main-sequence life-time 
of high-mass stars is performed by means of helium abundances.  
Abundances of several light elements were also obtained for a zero-age 
main-sequence eclipsing binary in {\sc ngc}\,2244, profiting from a precise 
determination of the gravity and the temperature ratio between the components, 
and thus leaving less ambiguity in the chemical composition. Note that 
several systems mentioned in Table~\ref{tab:lit} have components with a 
peculiar atmospheric composition, some of them revealing their peculiarity 
only after the spectra were disentangled. The first direct determination of 
the mass of a BpSi-type star was performed in AO\,Vel, and this quadruple 
system deserves better than the limited data set studied at present. Among 
the multiple systems studied to provide clues to the inter-relations between 
pulsation, rotation, chemical peculiarities and binarity in the domain of 
intermediate-mass stars (around late-A spectral types), DG\,Leo consists of 
a close binary with two metallic-line stars and an equal-mass wide companion 
that is pulsating. 

Although pulsating stars, and especially line-profile variables, violate 
the basic assumptions, the technique has proven its usefulness. Several 
applications deal with $\beta$~Cep-type stars. The disturbance of the 
companion on the line-profile variations of the pulsating component can be 
removed in order to facilitate the identification of the pulsation modes and 
the assignment to a particular component (see also Aerts 2007). 
%%% in this symposium). 
The success of these studies is for part due to the large number 
of input spectra which de-correlated effectively the line-shape variability 
from the orbital phase, such that the procedure used to disentangle the 
spectra sees the variability merely as an extra ``random noise'' relative to 
orbital phase. This is not guaranteed, as pulsation periods and orbital 
periods, although very different, may by chance be aliases of each other.  
It is e.g.\ also untrue for line-profile changes in semi-detached systems, 
where the changes are phase-locked to the orbital cycle, which may lead to 
the detection of spurious components (Bak\i \c{s} \etal\ 2006). \cite{Hadrava2004} 
has described how to generalise the technique to disentangle spectra in order 
to include certain types of intrinsic stellar variability and how to probe the 
stellar atmosphere by analyzing spectra obtained in partial eclipses, 
especially in the presence of the Schlesinger-Rossiter effect. The development 
of such generalised algorithms would significantly broaden the range of systems 
to which the reconstruction techniques can be applied with high confidence. 

Several papers deal with the fundamental parameters of high-mass stars, 
some of them highly evolved, in the Cas OB6 region (incl. IC\,1805), 
in the Galactic Center (an extremely high-mass binary), and beyond our Galaxy. 
An important aim of studies of eclipsing binaries in other galaxies is 
to contribute to the calibration of the distance scale. The most extensive 
application since previous reviews was performed by \cite{Hilditchetal2005} 
in the Small Magellanic Cloud. Together with their previous work 
(\cite{Harriesetal2003}) 
they alltogether disentangled 50 eclipsing binaries. Their 
sample comprises detached, semi-detached, and contact binaries. 
\cite{Ribasetal2005}  and \cite{Bonanosetal2006} studied eclipsing binaries 
in M31 and M33, respectively. The low {\sc s/n} spectra, even while secured 
at the worlds largest telescopes, apparently hamper disentangling efforts 
(although \cite{Ribasetal2005} succeeded to separate the component spectra 
with fixed orbital parameters), but it might be worthwhile to investigate 
whether the limitation is due to too low {\sc s/n} or to bias in the input 
data (normalisation, blemishes at low light level, interstellar bands, etc).

In several of the analyses of the Ond\v{r}ejov group, the telluric lines are 
separated from the stellar components in Fourier space. The approximation 
is good as long as the telluric line does not move across a large stellar 
spectral gradient. \cite{Hadrava2006} showed that variability in the intensity 
of spectral lines can be used to disentangle one component from another, 
even in absence of Doppler shifts. He separated in this way telluric from 
stellar lines in a set of spectra obtained in a short time-interval. The same 
paper discusses an extension of the disentangling of spectra to include 
components with a known spectrum ({\it constrained disentangling}). Avoiding 
in this way the introduction of a large amount of parameters, by exploiting 
prior knowledge e.g.\ on the telluric line spectrum or on the interstellar 
spectrum, increases the robustness of the analysis. The first application of 
this concept is shown in \cite{Hadrava2007}.

\section{Bias in the reconstructed spectra}\label{sec:risk}

\subsection{Nearly-singular equations}\label{sub:det}

Depending on the data set, some of the equations may become (nearly)-singular. 
Insight can be gained from studying the case of a binary star in the algorithm 
using the Fourier components of the input spectra, as the singularity can be 
coupled directly to specific Fourier modes. The determinant $D$ of the set of 
equations for Fourier mode $m$ is, for $N$ bins in the observed spectra,  
$$
D^{\frac{1}{2}} = \sum_{k'=1}^{k-1} \sum_{k=2}^{K} \left(\ell_1 (\phi_k) 
- \ell_1 (\phi_{k'}) \right)^2 
+ 2 \sum_{k'=1}^{k-1} \sum_{k=2}^{K} \ell_1 (\phi_k) \ell_1 (\phi_{k'}) 
\ell_2 (\phi_k) \ell_2 (\phi_{k'}) \left(1 - \cos x\right) %%% \nonumber 
\eqno(3)
$$
with 
$ x = 2\upi\frac{m}{N}\left(v_2 (\phi_k) - v_1 (\phi_k) - v_2 (\phi_{k'}) + v_1
(\phi_{k'})\right) $ and $-\frac{N}{2} + 1 < m < \frac{N}{2}$. The square-root 
of $D$ is expressed as a sum of non-negative terms. Each term refers to a 
pair of orbital phases $\phi_{k}, \phi_{k'}$ and involves the corresponding 
light contributions $\ell_1$ and $\ell_2 = 1 - \ell_1$ and the relative 
Doppler shifts $v_2 - v_1$.

In the case of significant light variability, the first sum of terms 
guarantees that no singularities occur. The continuum level in the 
component spectra is then well-determined. In absence of light variability, 
the determinant is strictly 0 for $m = 0$ corresponding to the intrinsic 
uncertainty how to distribute the observed line blocking over the two 
components, as mentioned earlier (Sect. \ref{sub:inp}). Near-singularities 
then exist likely for other low-frequency modes ($m << N$), responsible for 
the undulations in the component spectra mentioned in various papers 
%%%KP  If you think this is not appropriate just delete, but also
%%%KP  delete citation in list of references
(e.g.\ Hensberge \etal\ 2000; \cite{Fitzpatricketal2003};  
\cite{PavlovskiHensberge2005}; Gonz\'{a}lez \& Levato 2006).
Often no attention is paid to the fact that the bias introduced in one 
component is strictly correlated with the bias in the other component (in 
antiphase and amplitude proportional to $\ell_j^{-1}$). Continuum windows 
in one component suffice to remove the bias in both component spectra. 

\begin{figure}
\center
\includegraphics[height=12cm,angle=270]{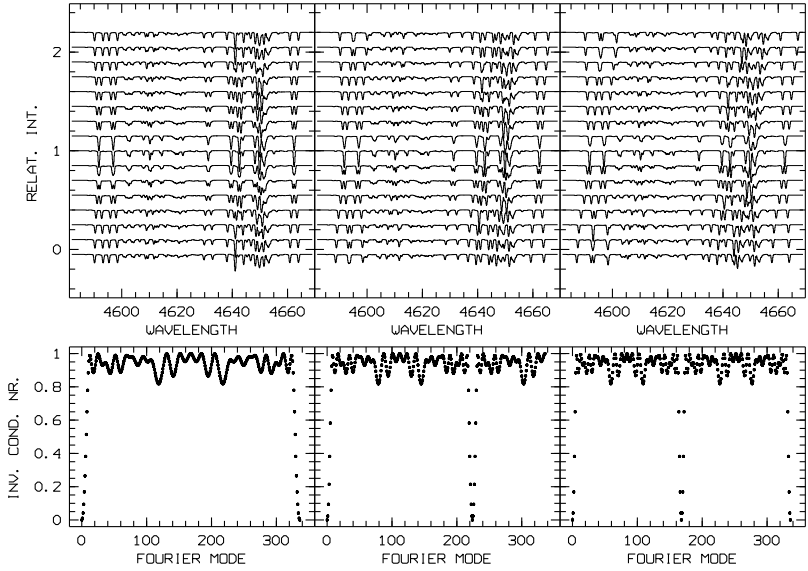}
\caption{Series of 16 artificial two-component spectra and corresponding
inverse condition numbers for non-negative Fourier modes. The relative Doppler
velocities, from left to right: $\frac{K2}{K1} = 1, 2, 3$ respectively, and the
orbital phases were chosen in order to reproduce cases with singular low-
and high-frequency Fourier modes (inverse condition number equal to zero)}
\label{fig:1}
\end{figure}

Numerical singularities appear in high-frequency modes when the argument of 
the cosine function is a multiple of $2 \upi$ for all (most) pairs of observed 
spectra, which occurs for integer values of 
$\frac{N}{m} \left(v_2 (\phi_k) - v_1 (\phi_k) - v_2 (\phi_{k'}) + v_1
(\phi_{k'})\right)$ (Fig.~\ref{fig:1}).  
The equations for (nearly)-singular modes should be solved using the 
singular value decomposition technique. On this condition, the singularity in 
high-frequency modes can be shown to be of practical concern only when $N$ 
is small i.e. when applying the method on single spectral lines, since the 
amplitude of the noise pattern is inversely proportional to $N$. 

The key point is that the occurrence of singularities depends in a predictable 
way on the distribution of the observations  over the orbit , on the level of 
time-variability in the relative light contributions, and on the chosen 
log-wavelength sampling.

\subsection{Biased input data}\label{sub:bias}

\begin{figure}[!ht]
\center
\includegraphics[height=10.5cm,angle=270]{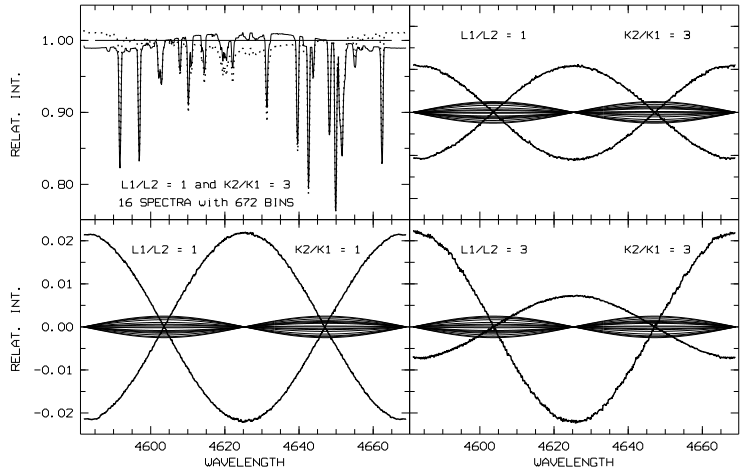}
\caption{Separation of component spectra applied the data set shown in the
rightmost panel of Fig.~\ref{fig:1} (upper panels) and similar ones (same
component spectra, but different light and velocity amplitude ratios, as
indicated in the panels. All spectra have $K1 = 12$\,bins). An orbital-phase
dependent bias was added to the observed spectra
$S_{obs}(\phi, \mathrm{ln}\lambda)$, i.e.\
$S_{obs} \rightarrow \frac{S_{obs}}{1 - 0.0025 \sin 2\upi\phi sin 2\upi\psi}$
with $\psi = \frac{\mathrm{ln}\lambda - \mathrm{ln}\lambda_{start}}
{\mathrm{ln}\lambda_{end} - \mathrm{ln}\lambda_{start}}$. Resulting separated
component spectra (left panel) show an amplified sinusoidal bias, of which a
detailed view is shown in the other panels for three different cases. The
set of 16 thin-line sine curves in these panels show the bias in the input
spectra (never and nowhere larger than 0.25\%), the two thick-line sine curves
indicate the amplified bias in the output spectra.}
\label{fig:2}
\end{figure}
 
\begin{figure}
\center
\includegraphics[height=11cm,angle=270]{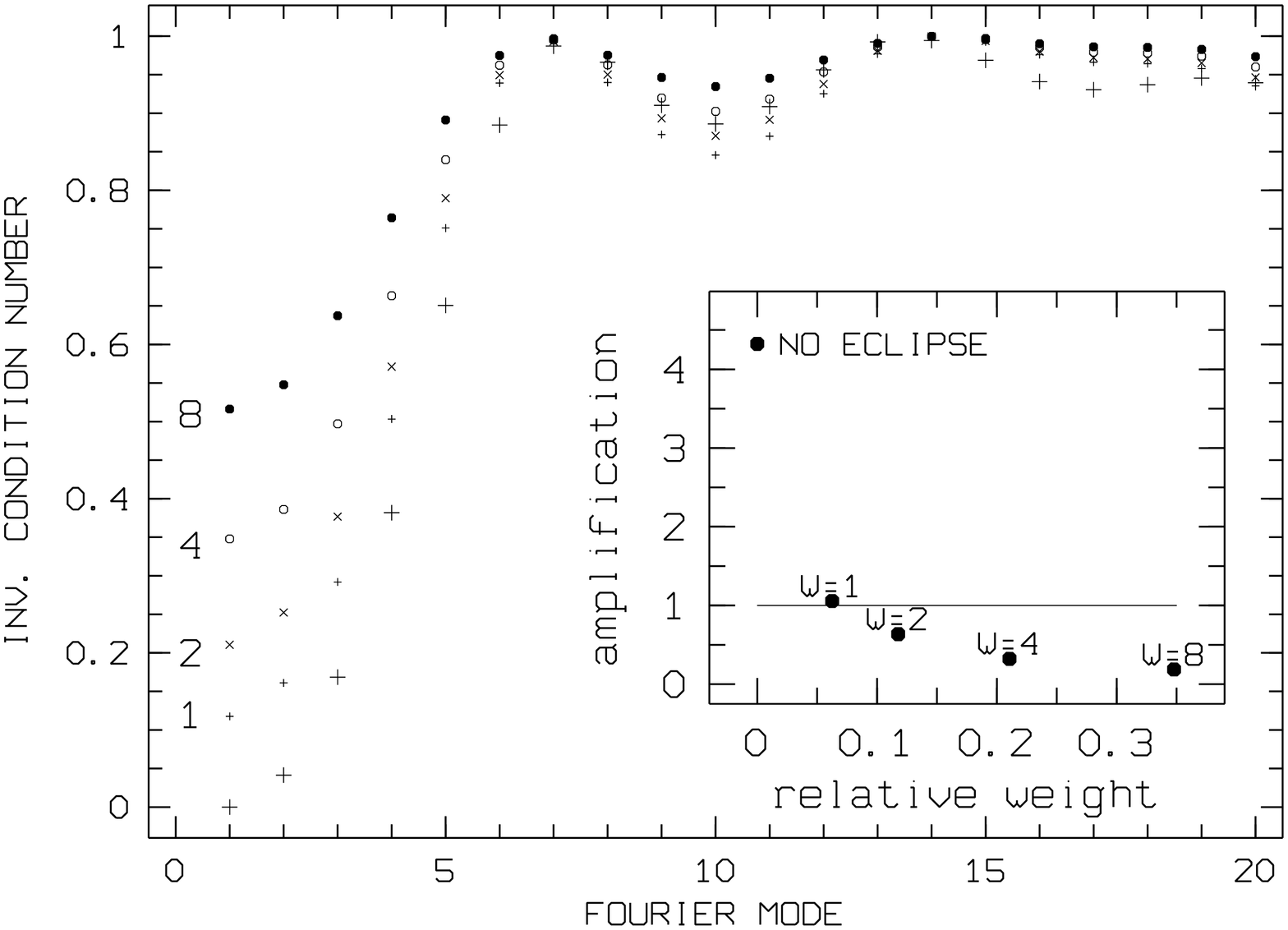}
\caption{Removal of the (near)-singularity of the equations in low Fourier
modes, and reduction of the amplitude of the bias in the component spectra
(small frame, ``amplification'' is the ratio of the amplitude of the bias in
the output spectra relative to the input spectra) in case one of the spectra
is taken in a total eclipse is shown for different weights $W$ given to the
eclipse spectrum.}
\label{fig:3}
\end{figure}

Multi-component spectra are often quite complex, because of the twice higher 
line density and the dilution of spectral lines. 
Especially in late-type spectra of close binaries, the synchronisation of the 
orbital motion and the stellar rotation may cause lines to be broader and 
shallower than in single stars. All these elements conspire to obscure the 
position of the continuum and the time-dependent Doppler shifts may lead to 
trace an observed (pseudo)-continuum that is biased with a dependence on 
orbital phase. How will the process of  separation of spectra react on such 
type of bias? 

Experiments with artificial data to which different types of phase-dependent 
bias was added show that the amplitude of the bias in the component spectra 
may be significantly larger than in the input spectra (Fig~\ref{fig:2}). The 
amplification is proportional to the ratio of the length of the spectral 
interval to the sum of the maximum Doppler shifts and inversely proportional 
to the relative light contribution $\ell_j$. However, mid-eclipse spectra 
reduce such low-frequency bias to a fraction of the bias in the input  
spectra when weights are applied in the low $m$ modes (Fig.~\ref{fig:3}). 
While the shape of the line profiles during eclipse might cast doubt about the 
usefulness of mid-eclipse spectra in the high-frequency modes, the advantages 
of their inclusion in the solution for low $m$ must be emphasized. 

Other types of bias encountered in observed spectra include shallow features, 
e.g. weak interstellar bands, detector blemishes or unidentified faint 
stellar components. Static features will either be included in a static 
stellar component and be amplified by $\ell^{-1}$, or, in the absence of such 
a component, they are at least slightly deformed and enter partially in the 
different components, inversely proportional as well to the stellar velocity
amplitude $K_j$ as to $\ell_j$. Such features not belonging to any of 
the components and undetected in the observed spectra are sometimes clearly 
recognized in one of the output spectra.  

The previous comments apply to separation of the spectra with known orbital 
parameters. The disentangling process is more complex, since any bias in the 
input will also influence the orbital parameters. \cite{HynesMaxted1998} 
discuss, based on numerical simulations, the relation between random noise in 
the input data and the uncertainty of the velocity amplitudes of the 
components. Also Iliji{\'c} showed, in the aforementioned meeting in 
Dubrovnik, that the uncertainty on the velocity estimates may be significantly 
too optimist when it is derived from a cross-correlation of the observed 
spectra with the disentangled component spectra {\it without} taking into 
account that the intensities in the component spectra were also parameters. 
There is indeed feed-back between residuals in the velocities and residuals 
in the component spectra. Realistic $\chi^2$--surfaces taking into account all 
parameters do not have the symmetry expected when the velocity amplitude 
estimates were independent of the errors in the reconstructed component 
spectra. The matter is relevant for the precision on the stellar masses. It 
relates also to the question in which conditions the reconstructed spectra do 
a better job, in terms of velocity amplitudes, than methods using 
``independent'' templates. Are there limits depending on {\sc s/n}, orbital 
coverage, richness of the line spectrum, \ldots? 

\begin{acknowledgments}
{\sc hh} acknowledges the project ``IUAP P5/36'' financed by the Belgian 
Science Policy. {\sc kp} acknowledges funding by the Croatian Ministry of 
Science under the project \#0119254. We thank Sasa Iliji\'{c} and Kelly 
Torres for contributions to and discussions on Section 5. 
\end{acknowledgments}

\end{document}